\PassOptionsToPackage{dvipsnames,svgnames,x11names}{xcolor}
\documentclass[10pt,a4paper,onecolumn]{article}
\usepackage{marginnote}
\usepackage{graphicx}
\usepackage{xcolor}
\usepackage{etoolbox}
\usepackage[affil-it]{authblk}
\usepackage{orcidlink}
\usepackage{titlesec}
\usepackage{calc}
\usepackage{tikz}
\usepackage{hyperref}
\usepackage{caption}
\usepackage{tcolorbox}
\usepackage{amssymb,amsmath}
\usepackage{ifxetex,ifluatex}
\usepackage{seqsplit}
\usepackage{xstring}

\usepackage{float}
\let\origfigure\figure
\let\endorigfigure\endfigure
\renewenvironment{figure}[1][2] {
    \expandafter\origfigure\expandafter[H]
} {
    \endorigfigure
}

\NewDocumentCommand\citeproctext{}{}
\NewDocumentCommand\citeproc{mm}{%
  \begingroup\def\citeproctext{#2}\cite{#1}\endgroup}
\makeatletter
 \let\@cite@ofmt\@firstofone
 \def\@biblabel#1{}
 \def\@cite#1#2{{#1\if@tempswa , #2\fi}}
\makeatother
\newlength{\cslhangindent}
\newlength{\csllabelwidth}
\newenvironment{CSLReferences}[2] % #1 hanging-indent, #2 entry-spacing
 {\begin{list}{}{%
  \setlength{\itemindent}{0pt}
  \setlength{\leftmargin}{0pt}
  \setlength{\parsep}{0pt}
  \ifodd #1
   \setlength{\leftmargin}{\cslhangindent}
   \setlength{\itemindent}{-1\cslhangindent}
  \fi
  \setlength{\itemsep}{#2\baselineskip}}}
 {\end{list}}
\usepackage{calc}

\ifLuaTeX
\usepackage[bidi=basic]{babel}
\else
\usepackage[bidi=default]{babel}
\fi
\babelprovide[main,import]{american}

\let\textttOrig=\texttt
\def\texttt#1{\expandafter\textttOrig{\seqsplit{#1}}}
\renewcommand{\seqinsert}{\ifmmode
  \allowbreak
  \else\penalty6000\hspace{0pt plus 0.02em}\fi}

\makeatletter
\let\href@Orig=\href
\def\href@Urllike#1#2{\href@Orig{#1}{\begingroup
    \def\Url@String{#2}\Url@FormatString
    \endgroup}}
\def\href@Notdoi#1#2{\def\tempa{#1}\def\tempb{#2}%
  \ifx\tempa\tempb\relax\href@Urllike{#1}{#2}\else
  \href@Orig{#1}{#2}\fi}
\def\href#1#2{%
  \IfBeginWith{#1}{https://doi.org}%
  {\href@Urllike{#1}{#2}}{\href@Notdoi{#1}{#2}}}
\makeatother

\usepackage[top=3.5cm, bottom=3cm, right=1.5cm, left=1.0cm,
            headheight=2.2cm, reversemp, includemp, marginparwidth=4.5cm]{geometry}

\titleformat{\section}
  {\normalfont\sffamily\Large\bfseries}
  {}{0pt}{}
\titleformat{\subsection}
  {\normalfont\sffamily\large\bfseries}
  {}{0pt}{}
\titleformat{\subsubsection}
  {\normalfont\sffamily\bfseries}
  {}{0pt}{}
\titleformat*{\paragraph}
  {\sffamily\normalsize}

\usepackage{fancyhdr}
\makeatletter
\let\ps@plain\ps@fancy
\fancyheadoffset[L]{4.5cm}
\fancyfootoffset[L]{4.5cm}

\definecolor{linky}{rgb}{0.0, 0.5, 1.0}

\newtcolorbox{repobox}
   {colback=red, colframe=red!75!black,
     boxrule=0.5pt, arc=2pt, left=6pt, right=6pt, top=3pt, bottom=3pt}

\newcommand{\ExternalLink}{%
   \tikz[x=1.2ex, y=1.2ex, baseline=-0.05ex]{%
       \begin{scope}[x=1ex, y=1ex]
           \clip (-0.1,-0.1)
               --++ (-0, 1.2)
               --++ (0.6, 0)
               --++ (0, -0.6)
               --++ (0.6, 0)
               --++ (0, -1);
           \path[draw,
               line width = 0.5,
               rounded corners=0.5]
               (0,0) rectangle (1,1);
       \end{scope}
       \path[draw, line width = 0.5] (0.5, 0.5)
           -- (1, 1);
       \path[draw, line width = 0.5] (0.6, 1)
           -- (1, 1) -- (1, 0.6);
       }
   }

\patchcmd{\@maketitle}{center}{flushleft}{}{}
\patchcmd{\@maketitle}{center}{flushleft}{}{}
\patchcmd{\@maketitle}{\LARGE}{\LARGE\sffamily}{}{}
\def\maketitle{{%
  
  \AB@maketitle}}
\makeatletter
\renewcommand\AB@affilsepx{ \protect\Affilfont}
\renewcommand\AB@affilnote[1]{{\bfseries #1}\hspace{3pt}}
\renewcommand{\affil}[2][]%
   {\newaffiltrue\let\AB@blk@and\AB@pand
      \if\relax#1\relax\def\AB@note{\AB@thenote}\else\def\AB@note{#1}%
        \setcounter{Maxaffil}{0}\fi
        \begingroup
        \let\href=\href@Orig
        \let\texttt=\textttOrig
        \let\protect\@unexpandable@protect
        \def\thanks{\protect\thanks}\def\footnote{\protect\footnote}%
        \@temptokena=\expandafter{\AB@authors}%
        {\def\\{\protect\\\protect\Affilfont}\xdef\AB@temp{#2}}%
         \xdef\AB@authors{\the\@temptokena\AB@las\AB@au@str
         \protect\\[\affilsep]\protect\Affilfont\AB@temp}%
         \gdef\AB@las{}\gdef\AB@au@str{}%
        {\def\\{, \ignorespaces}\xdef\AB@temp{#2}}%
        \@temptokena=\expandafter{\AB@affillist}%
        \xdef\AB@affillist{\the\@temptokena \AB@affilsep
          \AB@affilnote{\AB@note}\protect\Affilfont\AB@temp}%
      \endgroup
       \let\AB@affilsep\AB@affilsepx
}
\makeatother

\renewcommand\Affilfont{\sffamily\small\mdseries}
\ifnum 0\ifxetex 1\fi\ifluatex 1\fi=0 % if pdftex
  \usepackage[T1]{fontenc}
  \usepackage[utf8]{inputenc}

\else % if luatex or xelatex
  \ifxetex
    \usepackage{mathspec}
    \usepackage{fontspec}

  \else
    \usepackage{fontspec}
  \fi
  \defaultfontfeatures{Ligatures=TeX,Scale=MatchLowercase}

\fi
\IfFileExists{upquote.sty}{\usepackage{upquote}}{}
\IfFileExists{microtype.sty}{%
\usepackage{microtype}
\UseMicrotypeSet[protrusion]{basicmath} % disable protrusion for tt fonts
}{}

\usepackage{hyperref}
\hypersetup{unicode=true,
            pdftitle={\texttt{squishyplanet}: modeling transits of non-spherical exoplanets in JAX},
            pdfborder={0 0 0},
            breaklinks=true,
            pdfauthor={Ben Cassese, Justin Vega, Tiger Lu, Malena Rice, Avishi Poddar, David Kipping},
            pdflang={en-US},
            colorlinks=true,
            linkcolor={Maroon},
            filecolor={Maroon},
            citecolor=linky,
            urlcolor=linky,
            colorlinks,breaklinks=true,
            urlcolor=[rgb]{0.0, 0.5, 1.0},
            linkcolor=[rgb]{0.0, 0.5, 1.0}}
\let\addcontentslineOrig=\addcontentsline
\def\addcontentsline#1#2#3{\bgroup
  \let\texttt=\textttOrig\addcontentslineOrig{#1}{#2}{#3}\egroup}
\let\markbothOrig\markboth
\def\markboth#1#2{\bgroup
  \let\texttt=\textttOrig\markbothOrig{#1}{#2}\egroup}
\let\markrightOrig\markright
\def\markright#1{\bgroup
  \let\texttt=\textttOrig\markrightOrig{#1}\egroup}

\IfFileExists{parskip.sty}{%
\usepackage{parskip}
}{% else
\setlength{\parindent}{0pt}
\setlength{\parskip}{6pt plus 2pt minus 1pt}
}
\ifx\paragraph\undefined\else
\let\oldparagraph\paragraph
\renewcommand{\paragraph}[1]{\oldparagraph{#1}\mbox{}}
\fi
\ifx\subparagraph\undefined\else
\let\oldsubparagraph\subparagraph
\renewcommand{\subparagraph}[1]{\oldsubparagraph{#1}\mbox{}}
\fi

\title{\texttt{squishyplanet}: modeling transits of non-spherical exoplanets in JAX}

\author[1,2%
  \ensuremath\mathparagraph]{Ben Cassese%
    \,\orcidlink{0000-0002-9544-0118}\,%
    }
\author[1%
  ]{Justin Vega%
    \,\orcidlink{0000-0003-1481-8076}\,%
    }
\author[2%
  ]{Tiger Lu%
    \,\orcidlink{0000-0003-0834-8645}\,%
    }
\author[2%
  ]{Malena Rice%
    \,\orcidlink{0000-0002-7670-670X}\,%
    }
\author[1%
  ]{Avishi Poddar%
    \,\orcidlink{0009-0000-5314-5770}\,%
    }
\author[1%
  ]{David Kipping%
    \,\orcidlink{0000-0002-4365-7366}\,%
    }

\affil[1]{Dept. of Astronomy, Columbia University, 550 W 120th Street,
New York NY 10027, USA}
\affil[2]{Dept. of Astronomy, Yale University, New Haven, CT 06511, USA}
\affil[$\mathparagraph$]{Corresponding author}

\begin{document}
\maketitle

\marginpar{

  \begin{flushleft}
  %\hrule
  \sffamily\small

  {\bfseries DOI:} \href{https://doi.org/10.21105/joss.06972}{\color{linky}{10.21105/joss.06972}}

  \vspace{2mm}

  {\bfseries Software}
  \begin{itemize}
    \setlength\itemsep{0em}
    \item \href{https://github.com/openjournals/joss-reviews/issues/6972}{\color{linky}{Review}} \ExternalLink
    \item \href{https://github.com/ben-cassese/squishyplanet}{\color{linky}{Repository}} \ExternalLink
    \item \href{https://doi.org/10.5281/zenodo.13377036}{\color{linky}{Archive}} \ExternalLink
  \end{itemize}

  \vspace{2mm}

  \par\noindent\hrulefill\par

  \vspace{2mm}

  {\bfseries Editor:} \href{https://mbobra.github.io/}{Monica Bobra} \ExternalLink \\
  \vspace{1mm}
    {\bfseries Reviewers:}
  \begin{itemize}
  \setlength\itemsep{0em}
    \item \href{https://github.com/rferrerc}{@rferrerc}
    \item \href{https://github.com/catrionamurray}{@catrionamurray}
    \end{itemize}
    \vspace{2mm}

  {\bfseries Submitted:} 09 May 2024\\
  {\bfseries Published:} 30 August 2024

  \vspace{2mm}
  {\bfseries License}\\
  Authors of papers retain copyright and release the work under a Creative Commons Attribution 4.0 International License (\href{http://creativecommons.org/licenses/by/4.0/}{\color{linky}{CC BY 4.0}}).

  \end{flushleft}
}

\vspace{-0.8cm}
\hypertarget{summary}{%
\section{Summary}\label{summary}}

While astronomers often assume that exoplanets are perfect spheres when analyzing observations, the subset of these distant worlds that are subject to strong tidal forces and/or rapid rotations are expected to be distinctly ellipsoidal or even triaxial. Since a planet's response to these forces is determined in part by its interior structure, measurements of an exoplanet's deviations from spherical symmetry can lead to powerful insights into its composition and surrounding environment. These shape deformations will imprint themselves on a planet's phase curve and transit lightcurve and cause small (1s-100s of parts per million) deviations from their spherical-planet counterparts. Until recently, these deviations were undetectable in typical real-world datasets due to limitations in photometric precision. Now, however, current and soon-to-come-online facilities such as JWST will routinely deliver observations that warrant the consideration of more complex models. To this end we present \texttt{squishyplanet}, a \texttt{JAX}-based Python package that implements an extension of the polynomial limb-darkened transit model presented in Agol et al. (\citeproc{ref-alfm}{2020}) to non-spherical (triaxial) planets, as well as routines for modeling reflection and emission phase curves.

\vspace{-0.3cm}
\hypertarget{statement-of-need}{%
\section{Statement of Need}\label{statement-of-need}}

The study of exoplanets, or planets that orbit stars beyond the sun, is a major focus of the astronomy community. Many of these studies center on the analysis of time series photometric (or spectroscopic) observations collected when a planet happens to pass through the line of sight between an observer and its host star. By modeling the fraction of starlight intercepted by the coincident planet, astronomers can deduce basic properties of the system such as the planet's relative size, its orbital period, and its orbital inclination.

The past 20 years have seen extensive work both on theoretical model development and computationally efficient implementations of these models. Notable examples include Mandel \& Agol (\citeproc{ref-mandel_agol}{2002}), Kreidberg (\citeproc{ref-batman}{2015}), and Foreman-Mackey et al. (\citeproc{ref-exoplanet}{2021}), though many other examples can be found. Though each of these packages make different choices, the majority of them (with notable exceptions, including Maxted (\citeproc{ref-ellc}{2016})\footnote{
Though \texttt{ellc}, and   \texttt{squishyplanet} share the same goal of modeling transits of   non-spherical planets, they differ in a few key ways. First,   \texttt{ellc} requires users to select from a set of predefined limb   darkening laws, while \texttt{squishyplanet} allows for any law that   can be cast as a polynomial (e.g.~high-order approximations to   grid-based models). Second, \texttt{ellc} allows for gravity-deformed   stars, while \texttt{squishyplanet} always models the central star as   a sphere and restricts triaxial deformations to the planet only.   Third, \texttt{ellc} allows users to model radial velocity curves,   including the Rossiter-McLaughlin effect, while \texttt{squishyplanet}   is focused on lightcurve modeling only. In terms of implementation,   \texttt{ellc} is written in Fortran and wrapped in Python, while   \texttt{squishyplanet} is written in Python/\texttt{JAX}. Also,   \texttt{ellc} integrates the flux blocked by the planet via 2D   numerical integration, while \texttt{squishyplanet} uses a 1D   numerical integration scheme. We believe that these tools will be   complementary and that users will benefit from having both available.}
) do share one common assumption: the planet under examination is a perfect sphere.

This is both a reasonable and immensely practical assumption. It is reasonable because firstly, a substantial fraction of planets, especially rocky planets, are likely quite close to perfect spheres (Earth's equatorial radius is only 43 km greater than its polar radius, a difference of 0.3\%). Secondly, at the precision of most survey datasets (e.g.~\emph{Kepler} and \emph{TESS}), even substantially flattened planets would be nearly indistinguishable from a spherical planet with the same on-sky projected area (\citeproc{ref-zhu2014}{Zhu et al., 2014}). It is practical since, somewhat miraculously, this assumption enables an analytic solution for the amount of flux blocked by the planet at each timestep. This is true even if the intensity of the stellar surface varies radially according to a nearly arbitrarily complex polynomial (\citeproc{ref-alfm}{Agol et al., 2020}).

However, for a small but growing number of datasets and targets, the reasonableness of this assumption will break down and lead to biased results. Many gas giant planets, in particular, are expected to be distinctly oblate or triaxial, either due to the effects of tidal deformation or rapid rotation (\citeproc{ref-barnes2003}{Barnes \& Fortney, 2003}). Looking within our own solar system, Jupiter and Saturn have oblateness values of roughly 0.06 and 0.1, respectively, due to their fast spins.

\begin{figure}
\centering
\includegraphics[width=\textwidth]{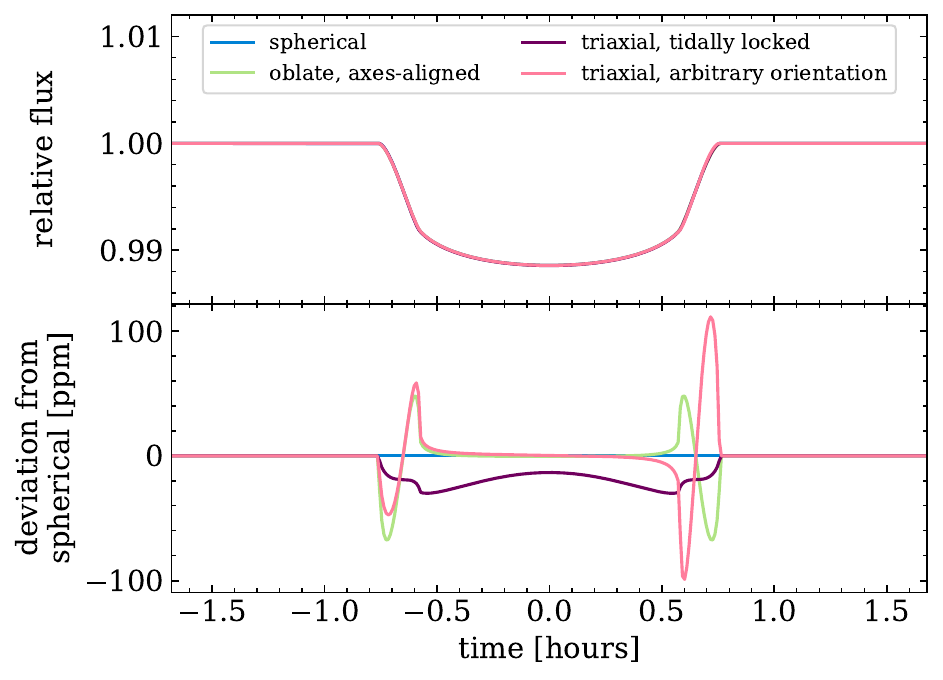}
\caption{A sampling of differences between transits of spherical and non-spherical planets. A more complete description of how each of these curves were generated can be found in the \protect\href{https://github.com/ben-cassese/squishyplanet/blob/main/joss/figure.py}{online documentation}.
\label{fig:example}}
\end{figure}

To illustrate the effects of shape deformation on a lightcurve, consider \autoref{fig:example}, which shows a selection of differences between time series generated under the assumption of a spherical planet and those generated assuming a planet with Saturn-like flattening. Depending on the obliquity, precession, impact parameter, and whether the planet is tidally locked, we can generate a wide variety of residual lightcurves. In some cases the deviations from a spherical planet occur almost exclusively in the ingress and egress phases of the transit, while others evolve throughout the transit. Some residual curves are mirrored about the transit midpoint, though in general, they will not always be symmetric (\citeproc{ref-carter_winn_empirical}{Carter \& Winn, 2010}).

\vspace{0.3cm}

The amplitudes of these effects are quite small compared to the full depth of the transit, but could be detectable with a facility such as JWST, which is capable of a white-light precision of a few 10s of ppm (\citeproc{ref-ERS_prism}{Rustamkulov et al., 2023}).

We leave a detailed description of the mathematics and a corresponding series of visualizations for the online documentation. There we also include confirmation that our implementation, when modeling the limiting case of a spherical planet, agrees with previous well-tested models even for high-order polynomial limb darkening laws. More specifically, we show that that lightcurves of spherical planets generated with \texttt{squishyplanet} deviate by no more than 100 ppb from those generated with \texttt{jaxoplanet} (\citeproc{ref-jaxoplanet}{Hattori et al., 2024}), the \texttt{JAX}-based rewrite of the popular transit modeling package \texttt{exoplanet} (\citeproc{ref-exoplanet}{Foreman-Mackey et al., 2021}) that also implements the arbitrary-order polynomial limb darkening algorithm presented in Agol et al. (\citeproc{ref-alfm}{2020}). Finally, we demonstrate \texttt{squishyplanet}'s limited support for phase curve modeling.

We hope that a publicly-available, well-documented, and highly accurate model for non-spherical transiting exoplanets will enable thorough studies of planets' shapes and lead to more data-informed constraints on their interior structures.

\hypertarget{acknowledgements}{%
\section{Acknowledgements}\label{acknowledgements}}

\texttt{squishyplanet} relies on \texttt{quadax} (\citeproc{ref-quadax}{Conlin, 2024}), an open-source library for numerical quadrature and integration in \texttt{JAX}. \texttt{squishyplanet} also uses the Kepler's equation solver from \texttt{jaxoplanet} (\citeproc{ref-jaxoplanet}{Hattori et al., 2024}) and the finite exposure time correction from \texttt{starry} (\citeproc{ref-starry}{Luger et al., 2019}). \texttt{squishyplanet} is built with the \texttt{JAX} library (\citeproc{ref-jax}{Bradbury et al., 2018}). We thank the developers of these packages for their work and for making their code available to the community.

\section*{References}\label{references}
\addcontentsline{toc}{section}{References}

\phantomsection\label{refs}

\end{document}